\documentclass[12pt, draftclsnofoot, onecolumn]{IEEEtran}
\usepackage{amsthm}
\usepackage{amsfonts}
\usepackage[top=0.8in, bottom=1in, left=0.7in, right=0.7in]{geometry}
\usepackage{amssymb}
\ifCLASSINFOpdf
  \usepackage[pdftex]{graphicx}
  \usepackage{graphicx}
  \graphicspath{{../pdf/}{../jpeg/}{../png/}}
  \DeclareGraphicsExtensions{.pdf,.jpeg,.png}
\else
  \usepackage[dvips]{graphicx}
  \graphicspath{{../eps/}}
  \DeclareGraphicsExtensions{.eps}
\fi
\ifCLASSOPTIONcompsoc
  \usepackage[caption=false,font=normalsize,labelfont=sf,textfont=sf]{subfig}
\else
\usepackage[caption=false,font=footnotesize]{subfig}
\fi
\usepackage{multirow}
\usepackage{balance}
\usepackage{multicol}
\usepackage{epstopdf}
\usepackage{verbatim} 

\usepackage[english]{babel}
\usepackage{float}
\usepackage{xcolor}
\usepackage[linesnumbered,ruled,vlined]{algorithm2e}

\usepackage[hidelinks]{hyperref}
\usepackage{cite}
\ifCLASSINFOpdf
   \usepackage[pdftex]{graphicx}
   \else
\usepackage[dvips]{graphicx}
\fi
\usepackage[cmex10]{amsmath}
\usepackage[T1]{fontenc}
\newcommand*{\affmark}[1][*]{\textsuperscript{\dag}}
\begin{document}

\title{QoS-Aware Energy-Efficient Sub-THz Near-Field Multiuser Hybrid MIMO Systems Employing Circular Arrays}
\author{
\normalsize{Priya~Gupta, Prabhanshu~Yadav, Suraj~Srivastava,  \\ Department of Electronics Engineering, Indian Institute of Technology Jodhpur, India\\(e-mail:  priyagupta@iitj.ac.in; p24ee0017@iitj.ac.in; surajsri@iitj.ac.in)}}
\maketitle
\thispagestyle{empty} 
\pagestyle{empty} 
\begin{abstract}
Near-field communications have emerged as a promising technology for future sixth-generation (6G) networks. However, designing energy-efficient hybrid beamfocusing schemes for multiuser (MU) systems remains challenging due to the non-convex quality-of-service (QoS) constraints and hardware limitations of hybrid architectures. In this paper, we investigate a hybrid massive multiple-input multiple-output (MIMO) architecture for a near-field sub-THz MU downlink system employing a large uniform circular array (UCA) at the base station. A QoS-aware transmit power minimization problem is formulated for energy-efficient beamfocusing and addressed through a two-stage penalty-driven joint optimization (PDJO) framework, where auxiliary variables are introduced to reformulate the QoS constraints into second-order cone (SOC) constraints. Furthermore, a near-field polar-domain codebook is developed via joint angle- and distance-domain sampling. We establish that the fully digital (FD) beamfocuser lies within the column space of the proposed codebook, enabling hybrid precoder design to be formulated as a multiple measurement vector (MMV)-based sparse approximation problem. The formulated problem is then solved using a low-complexity simultaneous orthogonal matching pursuit (SOMP) algorithm under hardware and transmit power constraints. Simulation results demonstrate that the proposed framework achieves spectral- and energy-efficiency performance close to the ideal FD beamfocusing scheme while providing a favorable trade-off between hardware complexity and system performance.
\end{abstract}
\begin{IEEEkeywords}
Uniform circular array, near-field, polar-domain codebook, hybrid beamfocusing, energy-efficiency, quality of service (QoS).
\end{IEEEkeywords}
\section{Introduction}\label{int_}
Sixth-generation (6G) wireless networks are envisioned to support a wide range of data-intensive and latency-sensitive applications, including holographic communications, immersive virtual and augmented reality (VR/AR), autonomous systems, and digital twin platforms~\cite{sarieddeen2021overview,prabhu_TGCN_1_Green_and_Safe}. Meeting the stringent requirements of these emerging services necessitates the exploitation of wider bandwidths beyond millimeter-wave (mmWave) frequency bands. In this context, sub-terahertz (sub-THz) communication, operating in the $100$--$300$ GHz spectrum, has emerged as a promising technology owing to its abundant spectrum resources, enabling ultra-high data rates with low transmission latency. Nevertheless, the practical deployment of sub-THz communication systems is challenged by severe propagation losses, susceptibility to signal blockage, and increased hardware complexity~\cite{akyildiz2022terahertz, heath2016overview, yong2006overview_intro,prabhu_TGCN_2_Spatial_Multiplexing_NF}. Leveraging the short wavelength at sub-THz frequencies, massive multiple-input multiple-output (MIMO) systems can accommodate a large number of antenna elements within a compact physical footprint, thereby providing substantial beamforming gains to compensate for the high path loss while simultaneously improving the signal-to-noise ratio (SNR) and spectral efficiency (SE)~\cite{prabhu_TGCN_3_Full_Duplex,prabhu_TGCN_4_First_20_years_Green_Radios}. However, implementing fully digital (FD) beamforming in massive MIMO systems requires a dedicated radio frequency (RF) chain per antenna element, resulting in prohibitive hardware complexity and system power consumption. Reducing the system power consumption of such large-scale antenna systems has become imperative for realizing green and sustainable 6G communications. Consequently, hybrid beamforming architectures have attracted significant attention as a practical alternative by employing substantially fewer RF chains than transmit/receive antennas while retaining most of the beamforming capabilities of FD architectures.

Furthermore, the adoption of extremely large antenna arrays at sub-THz frequencies leads to the transition from the conventional far-field region to the near-field regime, where the spherical nature of electromagnetic wave propagation can no longer be neglected. Consequently, conventional angle-only beam steering becomes insufficient, and beamforming must jointly exploit both angular and distance-domain information to accurately focus the electromagnetic beam toward the intended users, commonly referred to as beamfocusing. Although uniform linear arrays (ULAs) have been widely adopted in sub-THz MIMO systems due to their simple architecture, their applicability to near-field massive MIMO deployments remains limited~\cite{Zhangintro2}. Specifically, the effective aperture of a ULA decreases significantly for users located at large angles of incidence or departure, resulting in a considerable reduction in beamfocusing gain toward edge directions. As a consequence, the near-field coverage becomes highly angle dependent, leading to a non-uniform service region and degraded communication performance for users located away from the array broadside. Hence, the one-dimensional geometry of ULAs inherently restricts angular coverage, making them less suitable for wide-angle multiuser (MU) communications envisioned in future sub-THz networks.

Although uniform planar arrays (UPAs) improve spatial utilization and provide beamfocusing capability in both the horizontal and vertical dimensions, their near-field coverage remains angle dependent due to variations in the effective aperture. In contrast, uniform circular arrays (UCAs) maintain a nearly constant effective aperture over all azimuth directions, resulting in an almost angle-independent near-field region. Consequently, UCAs alleviate near-field contraction at wider angles, provide enhanced uniform spatial coverage, and enable more effective beamfocusing~\cite{yang2025near}. However, fully exploiting these advantages requires the careful design of hybrid precoders and combiners, which play a crucial role in determining the spectral and energy efficiency of sub-THz near-field MU-MIMO systems. Motivated by these observations, the following discussion reviews the existing literature on hybrid beamforming for near-field communications.












\subsection{Literature Review}
The design of efficient hybrid beamforming techniques for far-field MIMO systems has been extensively investigated to improve spectral efficiency (SE), satisfy quality-of-service (QoS) requirements, and enhance energy efficiency (EE) while mitigating the severe propagation loss encountered at sub-THz frequencies. In this direction, Wen \textit{et al.}~\cite{wen2023qos} proposed a user-centric QoS-aware beamforming framework for MU systems that minimizes the total transmit power while guaranteeing minimum rate requirements. However, the proposed design assumes far-field propagation and is developed for UPA-based architectures. Beyond QoS provisioning, energy-efficient hybrid beamforming strategies have also been investigated in~\cite{xue2020energy}, where iterative algorithms are employed to jointly optimize power allocation and beamforming vectors. Nevertheless, these methods are likewise established under the far-field assumption. More recently, Wu and Dai~\cite{wu2024near} examined the beam-defocus effect caused by the spatial wideband phenomenon in UCA-based wideband MIMO systems and proposed a delay-phase-precoding (DPP) architecture to mitigate this impairment. Although this work incorporates UCAs, its design is still tailored to far-field beamforming, leaving the unique beamfocusing challenges arising from spherical wavefront propagation in near-field communications largely unaddressed.

Recent studies on near-field communications have focused on developing beamfocusing techniques that explicitly account for spherical wavefront propagation. In this direction, Lu \textit{et al.}~\cite{lu2021near} investigated near-field modeling and performance analysis for extremely large-scale MU-MIMO systems by evaluating conventional beamfocusing schemes, including maximal-ratio combining (MRC), zero-forcing (ZF), and minimum mean-square error (MMSE). Furthermore, Zhang \textit{et al.}~\cite{zhang2022beam} studied near-field beamfocusing for high-rate MU downlink MIMO systems under FD and hybrid transmit architectures using a sum-rate maximization framework. However, these studies neither consider QoS-aware beamformer design nor extend beyond conventional ULA configurations.

With the emergence of wideband sub-THz communications, mitigating beam-splitting caused by spatial-wideband effects has become an important research direction in near-field beamfocusing. In this context, Cui \textit{et al.}~\cite{cui2024near} investigated wideband beamfocusing for extremely large-scale ULA systems by considering near-field propagation and spatial-wideband effects, and developed an efficient beamforming strategy to suppress frequency-dependent beam distortion. Similarly, Ding \textit{et al.}~\cite{11123647} proposed a three-stage hybrid beamfocusing approximation algorithm that jointly optimizes analog and digital beamformers to mitigate the beam-splitting effect in wideband near-field communications. Along the same direction, Yang \textit{et al.}~\cite{yang202near} proposed near-field direct greedy hybrid beamforming (NF-HBF) for high-frequency large-scale MIMO systems, along with stepwise-individual and stepwise-joint greedy schemes to reduce the computational complexity of analog and digital beamformer design. More recently, studies on ULA- and UPA-based near-field beamfocusing have also highlighted the impact of distance-angle coupling, which introduces additional challenges in beamformer design~\cite{chen2025near,liu2025near}. Despite these advances, the existing studies are primarily developed for ULA- and UPA-based architectures. To enhance user coverage and improve spatial multiplexing performance, Wu \textit{et al.}~\cite{wu2023enabling} investigated the transition from ULA to UCA configurations for near-field beamfocusing, demonstrating the potential of circular arrays for overcoming the limitations of conventional linear array geometries.

To the best of our knowledge, near-field communications employing UCA configurations have received relatively limited attention in the literature. Existing studies on near-field beamfocusing primarily focus on hybrid beamforming for conventional ULA- and UPA-based array geometries, which exhibit limited wide-angle beamfocusing capability in near-field communications. Although UCAs have recently emerged as a promising alternative owing to their nearly uniform angular coverage and improved beamfocusing characteristics, their integration with hybrid beamforming introduces new challenges in precoder design and practical hardware implementation. Therefore, there is a need to develop efficient hybrid beamforming techniques for UCA-based near-field MU systems that jointly address the performance and implementation challenges of future sub-THz communications. The key contributions of this work are summarized as follows:

\begin{table*}
    \centering
\caption{\small Comparison of the salient contributions of the proposed work with respect to already existing literature} \label{tab:lit_rev}
\begin{tabular}{|l|c|c|c|c|c|c|c|c|c|c|c|}

    \hline

\textbf{Features}
&\cite{yang2025near}
& \cite{wen2023qos} 
&\cite{lu2021near}
 &\cite{11123647} &\cite{wu2023enabling}  & \cite{wu2024near} &\cite{cui2024near}  & \cite{xue2020energy}&\cite{yang202near} & \cite{zhang2022beam} &\textbf{This Paper} \\

\hline
Sub-THz/ THz Band 
&  &  &  &  &  &  & \checkmark &  &  &  & \checkmark \\

\hline
Uniform circular array (UCA) 
& \checkmark &  &  &  & \checkmark &  &  & \checkmark &  &  & \checkmark \\

\hline
Near-field 
& \checkmark &  & \checkmark & \checkmark & \checkmark & \checkmark & \checkmark & \checkmark & \checkmark & \checkmark & \checkmark \\

\hline
MU scenario 
&  & \checkmark & \checkmark & \checkmark &  &  &  &  &  & \checkmark & \checkmark \\

\hline
Transmit power minimization 
&  & \checkmark &  &  &  &  &  &  & \checkmark &  & \checkmark \\

\hline
QoS constraints 
&  & \checkmark &  &  &  &  &  &  &  &  & \checkmark \\

\hline
{Hybrid beamformer design} 
&  &\checkmark  &  & \checkmark &  &  & \checkmark & \checkmark & \checkmark &  \checkmark & \checkmark \\

\hline
\end{tabular}\vspace{-1.5 \baselineskip}
\end{table*}


\subsection{Contributions}
\label{contri}
\begin{enumerate}
\item We formulate a QoS-aware energy-efficient beamfocusing design problem for a near-field sub-THz hybrid MU-MIMO downlink system employing a large UCA at the base station (BS). Specifically, the proposed optimization framework minimizes the total transmit power while guaranteeing the individual signal-to-interference-plus-noise ratio (SINR) requirements of all users. The resulting optimization problem is inherently non-convex due to the QoS constraints and constant-modulus constraints imposed on the RF precoder.

\item To solve the resulting non-convex optimization problem, we first introduce auxiliary variables to reformulate the QoS constraints into equivalent second-order cone (SOC) constraints, thereby enabling a more tractable optimization framework. Building upon this reformulation, we develop a two-stage penalty-driven joint optimization (PDJO) framework. In the first stage, a penalized version of the original problem is solved, whereas in the second stage, the penalty parameter is iteratively updated until convergence is achieved.

\item We develop a low-complexity hybrid transmit precoder design framework for the proposed near-field sub-THz MU-MIMO downlink system. To this end, a near-field polar-domain codebook is constructed through joint angle- and distance-domain sampling, where the sampling intervals are designed to ensure that the beamfocusing gain between the codebook vectors remains below a prescribed maximum correlation threshold.

\item We establish that the corresponding FD precoder lies in the column space of the underlying sub-THz near-field MU-MIMO downlink channel, which is further shown to be contained within the column space of the proposed polar-domain codebook. Building upon this theoretical result, we formulate a multiple measurement vector (MMV)-based sparse optimization problem that minimizes the approximation error between the FD and hybrid precoders while satisfying the associated hardware and transmit power constraints.

\item For the proposed MMV formulation, we develop a low-complexity hybrid precoder design framework using simultaneous orthogonal matching pursuit (SOMP). Requiring only a few iterations determined by the number of RF chains, the proposed scheme efficiently decomposes the FD precoder into its hybrid counterpart without requiring prior knowledge of the true near-field beamfocusing vectors.
 
\end{enumerate}

\subsection{Notation} \label{notation}
Upper-case letters $\mathbf{A}$ represent matrices, while lower-case letters $\mathbf{a}$ represent vectors. The quantity $\lVert \cdot \rVert_{F}$ represents the Frobenius norm and $\mathcal{CN}({\mu}, \sigma^2)$ represents a complex Gaussian distribution with mean ${\mu}$ and variance ${\sigma}^2$. Throughout this paper, $(\cdot)^H$ denotes the Hermitian (conjugate-transpose) operator. The operator $|\cdot|$ denotes the modulus (absolute value), while $\lVert \cdot \rVert$ denotes the Euclidean ($\ell_2$) norm. The operator $\lfloor \cdot \rfloor$ represents the floor function. The notation $\operatorname{diag}(\mathbf{A})$ denotes the diagonal matrix formed from the diagonal elements of the matrix $\mathbf{A}$. The notation $\mathcal{C}(\mathbf{A})$ denotes the column-space of the matrix $\mathbf{A}$.

\subsection{Paper organization}
The remainder of this paper is organized as follows: Section-\ref{2} presents the near-field sub-THz MU-MIMO system model, UCA-based channel model and QoS-aware transmit power minimization problem. Section-\ref{PD} introduces the proposed PDJO framework for FD MU-MIMO beamforming. Section-\ref{4} presents polar-domain near-field codebook development followed by near-field hybrid precoder design. Section-\ref{simulation} discusses our extensive simulation results, and finally Section-\ref{conclusion} concludes the paper.

\begin{figure*}
\centering
{\includegraphics[scale=0.55]{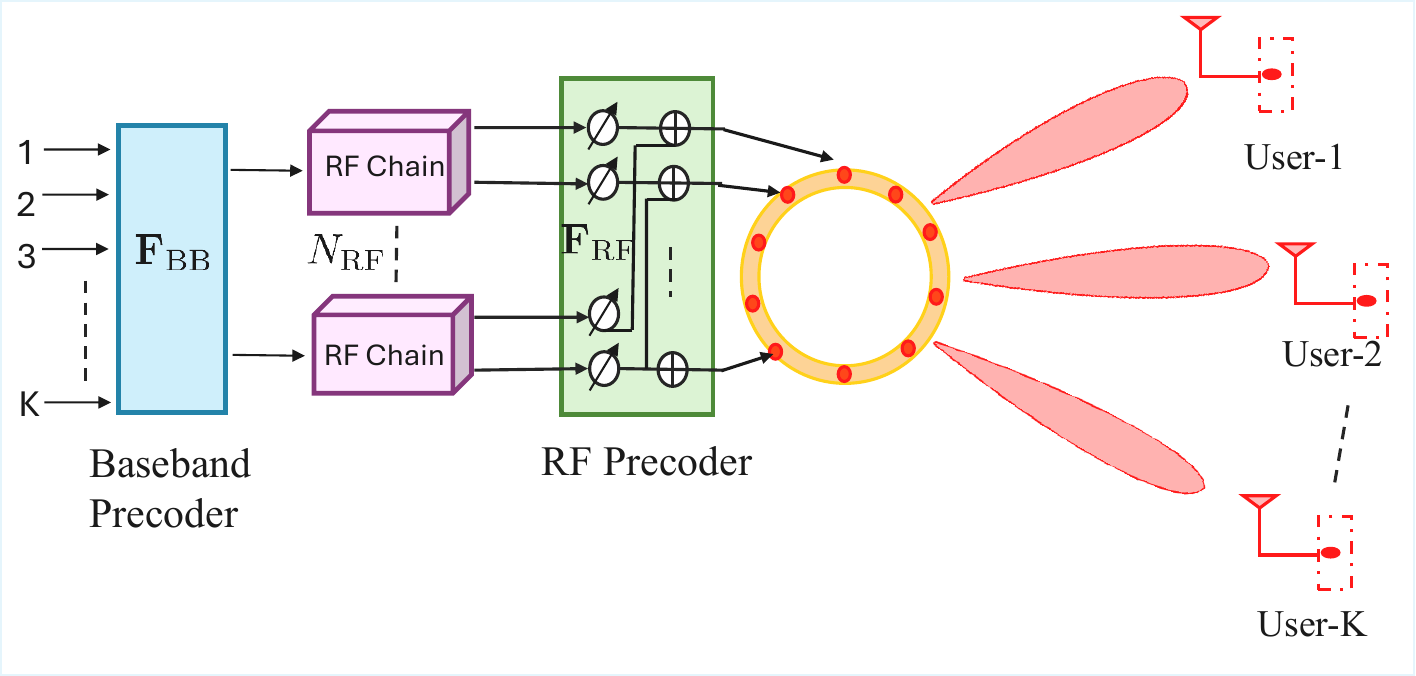}}
\caption{Near-field sub-THz MU downlink hybrid MIMO architecture relying on UCA-based BS}
\label{schematic}
\end{figure*}

\section{ Near-Field Sub-THz MU-MIMO System model }\label{2}
Consider a near-field sub-THz  hybrid MIMO architecture based MU downlink system, as depicted in Fig.~\ref{schematic}. The BS is equipped with a UCA employing $N$ transmit antennas and serving $K$ number of single-antenna users. The hybrid architecture at BS includes baseband precoder~$\mathbf{F}_{\mathrm{BB}}\in \mathbb{C}^{N_{\mathrm{RF}}\times K}$ implemented in digital signal processor, which is defined as $\mathbf{F}_{\mathrm{BB}}=[\mathbf{f}_{\mathrm{BB},1}, \mathbf{f}_{\mathrm{BB},2}, \cdots , \mathbf{f}_{\mathrm{BB},K}]$, where $\mathbf{f}_{\mathrm{BB},j}$ denotes baseband precoding vector for the $j$-th user and $N_{\mathrm{RF}}$ denotes the number of RF chains obeying $K \leq N_{\mathrm{RF}}<<N$. The RF precoder~$\mathbf{F}_{\mathrm{RF}}\in \mathbb{C}^{N \times N_{\mathrm{RF}}}$ is implemented in analog domain using a network of digitally controlled phase-shifters, designed under the constant modulus constraint as  
$|\mathbf{F}_{\mathrm{RF}}(i,j)| = \frac{1}{\sqrt{N}}$.

Let $s_{k}$ represent the $k$-th user's signal, which are assumed to be mutually independent across users and satisfy $\mathbb{E}\{|s_{k}|^2\}=1, \forall k \in \mathcal{K}\triangleq \{1,2,3, \cdots, K   \}$. 
For the downlink transmission, the BS precodes the $k$-th user's signal $s_{k}$ with its baseband precoder $\mathbf{f}_{\mathrm{BB},k}$ followed by RF precoding using $\mathbf{F}_{\mathrm{RF}}$. 
Subsequently, the BS constructs the superposed signal vector $\mathbf{x}\in \mathbb{C}^{N \times 1}$ as $$\mathbf{x}= \mathbf{F}_{\mathrm{RF}}\sum_{j=1}^{K}\mathbf{f}_{\mathrm{BB},j}s_{j}=\mathbf{F}_{\mathrm{RF}}\mathbf{F}_{\mathrm{BB}}\mathbf{s},$$ where $\mathbf{s} = \left[ s_1, s_2, \cdots, s_K \right]^T \in \mathbb{C}^{K \times 1}$, and transmits the vector $\mathbf{x}$  via its antennas.
The signal $y_{k} \in \mathbb{C}$ received at $k$-th user can be given as
\begin{align}\label{received_signal}
y_{k}=\mathbf{h}^{H}_{k}\mathbf{x}+n_{k},
\end{align}
where $\mathbf{h}_{k}\in \mathbb{C}^{N \times 1}$ represents a near-field channel from BS to $k$-th user, whose model is detailed in the subsequent subsection, and  $n_{k} \sim \mathcal{CN}(0,\sigma^2_{k})$ denotes additive white complex Gaussian noise~(AWGN) at the $k$-th user. 

\begin{figure}
\centering
{\includegraphics[scale=0.4]{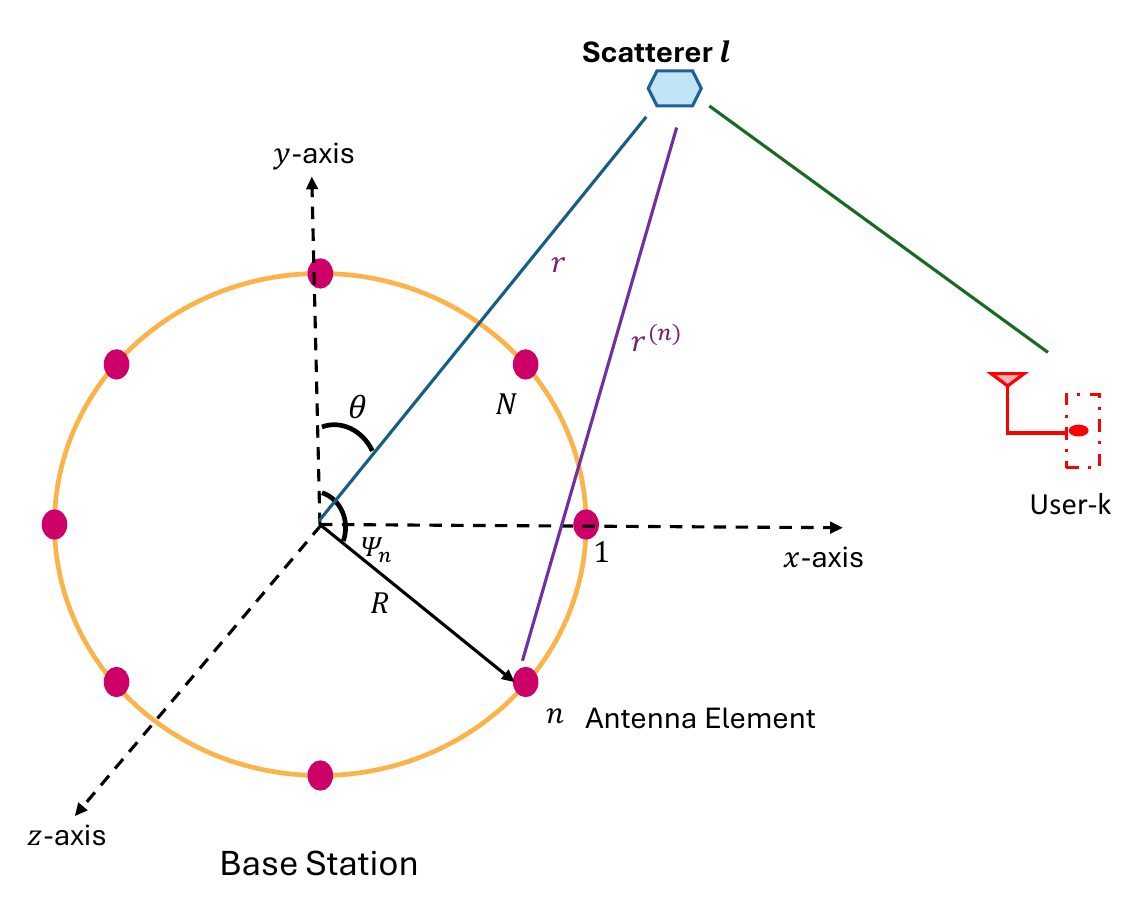}}
\caption{Geometric relation between UCA and $k$-th user in near-field with multipath propagation environment}
\label{UCA}
\end{figure}

\subsection{Near-Field UCA Channel Model} \label{channel_model}
Note that the typical far-field beam steering vector $\check{\mathbf{a}}(\theta) \in \mathbb{C}^{N \times 1}$ for a UCA is expressed as~\cite{wu2023enabling}
\begin{align}
    \!\!\!\!\!\check{\mathbf{a}}(\theta) =\frac{1}{\sqrt{N}}\left[e^{-j\frac{2 \pi}{\lambda}R\cos(\theta-\psi_1)}, \cdots, e^{-j\frac{2 \pi}{\lambda}R\cos(\theta-\psi_N)}\right]^T, \label{far_array}
\end{align}
where $R$ denotes the radius of the UCA on which the antennas are uniformly distributed, as illustrated in Fig~\ref{UCA}. The parameter $\psi_n$ represents the angular position of the 
$n$-th antenna element with respect to (w.r.t.) y-axis and it is defined as $\psi_n = \frac{2 \pi n}{N},\: \forall n = 1,2 ,\cdots, N$. Similarly, $\theta$ denotes the angle of departure (AoD) of the signal w.r.t. y-axis and $\lambda$ represents  carrier's wavelength. 
 
However, for sub-THz regime with large antenna arrays at the BS leading to near-field scenario, the above expression of beam steering vector, which assumes a far-field and planar wave propagation model, is no longer valid. Hence, for the scenario under consideration, the near-field beamfocusing vector $\tilde{\mathbf{a}}(r,\theta) \in \mathbb{C}^{N \times 1}$  characterizing the spherical wave propagation model is given by~\cite{zhang2019near}
\begin{align}
\tilde{\mathbf{a}}(r,\theta)
= \frac{1}{\sqrt{N}}\left[e^{-j \frac{2\pi}{\lambda}(r^{(1)}-r)},
\cdots, e^{-j \frac{2\pi}{\lambda}(r^{(N)}-r)}\right]^T, \label{eq:a_beam_focus}
\end{align}
where $r^{(n)}$ and $r$ denote the propagation distances from a scatterer to $n$-th antenna and to the center of UCA, respectively. The quantity $r^{(n)}$ can be mathematically computed as
\begin{equation}
    \begin{aligned}
         r^{(n)} &= \sqrt{r^{2} + R^{2} - 2rR \cos(\theta-\psi_{n})}\\  & \overset{(a)}{\approx} r - R \cos(\theta - \psi_{n}) + \frac{R^{2}}{2 r}[1 - \cos^{2}(\theta - \psi_{n})], \\
         & \approx r +\Xi^{(n)}_{r,\theta}.
    \end{aligned}\label{radius}
\end{equation}
Hence, the term $(r^{(n)}-r)$ appearing in \eqref{eq:a_beam_focus} can be approximated with $\Xi^{(n)}_{r,\theta} = - R \cos(\theta - \psi_{n}) + \frac{R^{2}}{2 r}[1 - \cos^{2}(\theta - \psi_{n})]$. Note that the above approximation is derived from the \textit{second-order Taylor series} expansion $\sqrt{1+x} = 1+\frac{x}{2}-\frac{x^2}{8}+\mathcal{O}(x^3)$, assuming $r$ to be very large compared to $R$, allowing higher-order terms to be neglected~\cite{wu2023enabling}. Finally, the near-field sub-THz MU-MIMO channel for $k$-th user can be modeled as \cite{wu2023enabling}
\begin{equation}\label{channel_model}
\mathbf{h}_{k}= \sum_{l=1}^{L}\alpha_{k,l}\,
\tilde{\mathbf{a}}(r_l,\theta_l),
\end{equation}
where $L$ represents the number of dominant scatterers and $\alpha_{k,l} \in \mathbb{C}$ denotes complex path gain corresponding to the $l$-th scatterer for $k$-th user. The quantities $r_l$ and $\theta_l$ represent distance and AoD, respectively, of the $l$-th scatterer, and the related beamfocusing vector $\tilde{\mathbf{a}}(r_l,\theta_l)$ is defined in \eqref{eq:a_beam_focus}. The QoS-aware total transmit power minimization problem is formulated in the next subsection.

\subsection{QoS-Aware Transmit Power Minimization Problem Formulation} 
We further rewrite Eq.~\eqref{received_signal} as follows
\begin{align}
y_{k}&=\mathbf{h}^{H}_{k} \mathbf{F}_{\mathrm{RF}}\sum_{j=1}^{K} \mathbf{f}_{\mathrm{BB},j}s_{j} +n_{k} \\ \nonumber
&=\underbrace{\mathbf{h}^{H}_{k}\mathbf{F}_{\mathrm{RF}}\mathbf{f}_{\mathrm{BB},k}s_{k}}_{\text{desired signal}}+\underbrace{\mathbf{h}^{H}_{k} \mathbf{F}_{\mathrm{RF}}\sum_{j\neq k} \mathbf{f}_{\mathrm{BB},j}s_{j}}_{\text{MU interference}} +n_{k}.
\end{align}
The $k$-th user's SINR, denoted by $\text{SINR}_{k}$, is given as
\begin{align}
\mathrm{SINR}_k = \frac{\left| \mathbf{h}_k^{H} \mathbf{F}_{\mathrm{RF}}\mathbf{f}_{\mathrm{BB},k} \right|^2}{\sum_{j \neq k} \left| \mathbf{h}_k^{H} \mathbf{F}_{\mathrm{RF}} \mathbf{f}_{\mathrm{BB},j} \right|^2 + \sigma_k^2}, \quad \forall\, k \in \mathcal{K}.
\end{align}
Furthermore, the total transmit power of BS is expressed as $$\mathbb{E}\{\parallel \mathbf{x} \parallel^2\} = \sum_{k=1}^{K} \|\mathbf{F}_{\mathrm{RF}}\mathbf{f}_{\mathrm{BB},k}\|^2 = \parallel \mathbf{F}_{\mathrm{RF}}\mathbf{F}_{\mathrm{BB}} \parallel^2.$$ 

We consider a QoS-constrained optimization problem that aims to minimize total transmit power at the BS through optimization of the transmit RF precoding matrix $\mathbf{F}_{\mathrm{RF}}$ and baseband precoders $\mathbf{f}_{\mathrm{BB},k}$. This optimization problem can be formulated as follows:
\begin{align}
\mathcal{P}_{0}: \min_{\{\mathbf{F}_{\mathrm{RF}}, \mathbf{F}_{\mathrm{BB}}\}} \quad & \parallel \mathbf{F}_{\mathrm{RF}}\mathbf{F}_{\mathrm{BB}} \parallel^2  \label{objective}
\\ \text{s.t.} \quad & \mathrm{SINR}_k \geq \gamma_k, \quad \forall k = 1, \ldots, \mathcal{K}, \label{SINR_constr}
\\ \quad & |\mathbf{F}_{\mathrm{RF}}(i,j)| = \frac{1}{\sqrt{N}}\label{Modu},
\end{align}
where $\gamma_{k}$ represents the minimum SINR requirement of $k$-th user. The problem defined in $\mathcal{P}_{0}$ is non-convex in nature due to the SINR constraint defined in~\eqref{SINR_constr} and the constant modulus constraints
associated with each element of the RF precoder $\mathbf{F}_{\mathrm{RF}}$ defined in~\eqref{Modu}. Solving the aforementioned non-convex optimization problem using conventional methods is highly challenging. To address this issue, we propose a two-step penalty-driven algorithm, which is detailed in the next section.

\section{Penalty-Driven Joint Optimization (PDJO) Framework for Near-Field MU-MIMO Beamfocusing}\label{PD}

In the proposed two-layer PDJO framework, Layer-1 employs an iterative approach to solve the resulting penalized optimization problem, while Layer-2 updates the penalty parameter until the overall convergence is achieved. To achieve this, we first define an equivalent FD beamformer for $j$-th user as  $\mathbf{w}_{j}=\mathbf{F}_{\mathrm{RF}}\mathbf{f}_{\mathrm{BB},j}$ and introduce the auxiliary variable~$c_{k,j}=\mathbf{h}_k^H\mathbf{w}_{j}$. Therefore, the non-convex constraint defined in~\eqref{SINR_constr} can be rewritten as
\begin{align}
&\frac{|c_{k,k}|^2}{\sum_{j \neq k}|c_{k,j}|^2+\sigma_{k}^2} \geq \gamma_{k},\  \forall k, \\
 & c_{k,j} = \mathbf{h}^H_{k}\mathbf{w}_{j}, \;\forall (k,j). \label{eq:c_k_j_def}
\end{align}
Thus, the PDJO problem corresponding to $\mathcal{P}_{0}$ of \eqref{objective} can be reformulated as follows:
\begin{align}\label{problem_1}
\!\!\!\mathcal{P}_{1}(\zeta) : \!\!\min_{\{\mathbf{w}_k\}, \{c_{k,j}\}} \quad 
& \!\!\!\!\!\!\sum_{k=1}^{K} \!\|\mathbf{w}_k\|^2 
\!+ \!\frac{\zeta}{2} \!\!\sum_{k=1}^{K} \sum_{j=1}^{K} 
\left| \mathbf{h}_k^H \mathbf{w}_j - c_{k,j} \right|^2 \\
\text{s.t.} \quad 
& \frac{|c_{k,k}|^2}{\sum\limits_{j \neq k} |c_{k,j}|^2 + \sigma_k^2} 
\geq \gamma_k, \quad \forall k, \label{SINR_2}
\end{align}
where $\zeta$ denotes the penalty factor. The first term $\sum_{k=1}^{K} \|\mathbf{w}_k\|^2$ denotes the total transmit power of BS, whereas the second term is associated with the auxiliary variables defined in \eqref{eq:c_k_j_def}. Furthermore, the constant modulus constraint of \eqref{Modu} is dropped in $\mathcal{P}_{1}(\zeta)$ above, since $\mathbf{w}_k$ is a FD beamformer. This constraint will be invoked later once again while designing the RF and baseband beamformers $\mathbf{F}_{\mathrm{RF}}$ and $\mathbf{f}_{\mathrm{BB},j}$, respectively. 
Note that the SINR constraint defined in~\eqref{SINR_2} is still non-convex, however it can be reformulated in the form of second-order cones as shown below
\begin{align}
\frac{|c_{k,k}|^2}{\gamma_k} + |c_{k,k}|^2 
&\geq \sum_{i \neq k} |c_{k,i}|^2 + \sigma_k^2 \nonumber \\
\implies |c_{k,k}|^2 \left(1 + \frac{1}{\gamma_k} \right) 
&\geq \sum_{i=1}^{K} |c_{k,i}|^2 + \sigma_k^2.
\end{align}
Let the elements of a matrix $\mathbf{D}\in \mathbb{C}^{K \times K}$ be defined as $\mathbf{D}[k,j]=c_{k,j}$, the above constraint can be  further reformulated as
\begin{align}
|c_{k,k}|^2 \left(1 + \frac{1}{\gamma_k} \right) 
&\geq \left\| \mathbf{D}^H \mathbf{e}_k \right\|_2^2 + \sigma_k^2 \nonumber\\
\implies \left(\sqrt{1 + \frac{1}{\gamma_k}}\right) \, c_{k,k} 
&\geq 
\left\|
\begin{array}{c}
\mathbf{D}^H \mathbf{e}_k \\
\sigma_k
\end{array}
\right\|_2, \quad \forall k,
\end{align} 
where the vector $\mathbf{e}_k \in \mathbb{R}^{K \times 1}$ denotes a canonical basis vector whose $k$-th element is equal to $1$, while all other elements are zero. Now the PDJO framework with the above reformulated constraints is expressed as
\begin{align}
\mathcal{P}_{2}(\zeta):\min_{\{{\mathbf{w}}_k\}, \{c_{k,j}\}} \quad 
& \sum_{k=1}^{K} \|{\mathbf{w}}_k\|^2 
+ \frac{\zeta}{2} \sum_{k=1}^{K} \sum_{j=1}^{K} 
\left| \mathbf{h}_k^H {\mathbf{w}}_j - c_{k,j} \right|^2 \\
\text{s.t.}&   \left(\sqrt{1 + \frac{1}{\gamma_k}} \,\right) c_{k,k} 
\geq 
\left\|
\begin{array}{c}
\mathbf{D}^H \mathbf{e}_k \\
\sigma_k
\end{array}
\right\|_2, \  \forall k. \label{CONE}
\end{align}
The reformulated problem $\mathcal{P}_{2}$ is non-convex due to coupling of optimization variables $\mathbf{w}_{k}$ and $c_{k,j}$, whereas the constraint specified in~\eqref{CONE} constitutes a second-order cone which is a convex set. In order to efficiently solve this overall non-convex problem, we adopt a two-stage solution as described next.

\begin{algorithm}[t]
\caption{Penalty-Driven Joint Optimization~(PDJO) for FD Beamformer Design}
\label{penalty_algorithm}
\KwIn{MU sub-THz MIMO channel $\mathbf{h}_k,  \forall k$  }
\KwOut{FD beamformer $\mathbf{W}$}

\textbf{Initialization:} $\zeta$ and  $c_{k,j}$

\While{$\Big(|\mathbf{h}^{H}_{k}\mathbf{w}_{j}-c_{k,j}|^2 > \epsilon$ \Big) }{

\Repeat{ (objective value of $\mathcal{P}_{2}(\zeta)$ converges within a tolerance limit $\delta$) }{

update $\mathbf{w}_k, \forall k,$ using \eqref{stage_1}

update $\{c_{k,j}\}, \forall (k,j),$ using \eqref{stage2_final} 
}

update $\zeta$ by using $\zeta=\frac{\zeta}{p}$, where $0<p<1$ 
}


\textbf{return:} $\mathbf{W}=\left[ \mathbf{w}_{1}, \mathbf{w}_{2},\cdots,\mathbf{w}_{K} \right]$
\end{algorithm}
\subsection{Stage-1}
For a given $c_{k,j}, \forall (k,j)$, the optimization problem $\mathcal{P}_{2}(\zeta)$ becomes unconstrained and it is given as
\begin{align}\label{opt}
\min_{\{{\mathbf{w}}_k\}} \quad 
& \sum_{k=1}^{K} \parallel{\mathbf{w}}_k\parallel^2 
+ \frac{\zeta}{2} \sum_{k=1}^{K} \sum_{j=1}^{K} 
\left| \mathbf{h}_k^H {\mathbf{w}}_j - c_{k,j} \right|^2. 
\end{align}
As derived in Appendix-\ref{PDJO_derivation}, computing  gradient of the above expression w.r.t. the beamformer $\mathbf{w}_{k}$ and setting the resultant expression to zero yields the solution as
\begin{align}\label{stage_1}
\mathbf{w}_{k}=\zeta \left[ 2\mathbf{I} + \zeta \sum_{j=1}^{K} \mathbf{h}_j \mathbf{h}_j^{H} \right]^{-1}\left(\sum_{j=1}^{K}\mathbf{h}_{j}c_{j,k}\right), \forall k.
\end{align}

\subsection{Stage-2}
For given $\mathbf{w}_{j}, \forall j\in \mathcal{K}$, obtained from stage-1, the problem in~\eqref{opt} becomes
\begin{align}\label{stage2_final}
\min_{\{c_{k,j}\}} \quad & \sum_{k=1}^{K} \sum_{j=1}^{K} \left| \mathbf{h}_k^{H} \mathbf{w}_j - c_{k,j} \right|^2 \\
\text{s.t.} \quad 
& \left(\sqrt{1 + \frac{1}{\gamma_k}} \,\right) c_{k,k} 
\geq 
\left\|
\begin{array}{c}
\mathbf{D}^H \mathbf{e}_k \\
\sigma_k
\end{array}
\right\|_2, \  \forall k. \nonumber
\end{align}
We observe that the above optimization problem defined in~\eqref{stage2_final} is convex w.r.t. the auxiliary variables $c_{k,j}$. Therefore, it can be efficiently solved using the convex-optimization solvers, such as CVX \cite{boyd2004convex}. Consequently, we obtain the solution for an equivalent FD beamformer $\mathbf{W}=[\mathbf{w}_{1}, \mathbf{w}_{2},\mathbf{w}_{3},\cdots,\mathbf{w}_{K}] \in \mathbb{C}^{N \times K}$.  

As discussed at the beginning of this section, the Layer-1 runs Stage-1 and Stage-2 iteratively till convergence for a given value of the penalty factor $\zeta$. Whereas in Layer-2, $\zeta$ is initially assigned with a small value to obtain a suitable starting point, and is subsequently increased in a gradual manner to progressively enforce the penalty constraint. This can be understood as follows: a smaller value of $\zeta$ in \eqref{problem_1} ensures transmit power minimization, which is the first part of the objective function; whereas a large value of $\zeta$ ensures $c_{k,j} = \mathbf{h}^H_{k}\mathbf{w}_{j}, \;\forall (k,j)$, which is the second part of the objective function. Hence, to start with a suitable solution that minimizes the transmit power, we initialize $\zeta$ with a very small value and update $\zeta=\frac{\zeta}{p}$, where the scalar $0<p<1$ is a constant. The PDJO framework for FD beamformer design in near-field sub-THz MU-MIMO system is presented in Algorithm~(\ref{penalty_algorithm}). The next section develops the design of hybrid precoder, which takes this FD beamformer $\mathbf{W}$ obtained from Algorithm~(\ref{penalty_algorithm}) as input, and outputs $\mathbf{F}_{\mathrm{RF}}$ and $\mathbf{F}_{\mathrm{BB}}$.

\section{Near-field Hybrid Precoder Design for Sub-THz MU-MIMO Systems}\label{4}
In order to design near-field hybrid precoder from the corresponding FD precoder $\mathbf{W}$, this section first develops an appropriate polar-domain codebook, followed by decomposing $\mathbf{W}$ into RF precoder $\mathbf{F}_{\mathrm{RF}}$ and baseband precoder $\mathbf{F}_{\mathrm{BB}}$.
\subsection{Polar-Domain Near-Field Codebook Design for sub-THz MIMO Systems}
We now begin with the beamfocusing gain $g(r_1,\theta_1,r_2,\theta_2)$ corresponding to the beamfocusing vectors $\tilde{\mathbf{a}}(r_1,\theta_1)$ and $\tilde{\mathbf{a}}(r_2,\theta_2)$, defined as 
    \begin{equation}
    \begin{aligned}
    g(r_1,\theta_1,r_2,\theta_2) &= \big|\tilde{\mathbf{a}}^H(r_1,\theta_1)\tilde{\mathbf{a}}(r_2,\theta_2)\big|\\
    &=\frac{1}{N} \bigg|\sum_{n=1}^{N} e^{j\frac{2\pi}{\lambda} (\Xi_{r_{1},\theta_{1}}^{(n)} - \Xi_{r_{2},\theta_{2}}^{(n)})} \bigg|\\ 
    &\approx \frac{1}{N} \left\vert \sum_{n=1}^N e^{j\frac{2 \pi}{\lambda}R(\cos(\theta_1-\psi_n)-\cos(\theta_2-\psi_n))} 
    \times  e^{-j\frac{2 \pi}{\lambda}R^2\Big(\frac{1-\cos^2(\theta_1-\psi_n)}{2 r_1}-\frac{1-\cos^2(\theta_2-\psi_n)}{2 r_2}\Big)} \right\vert. \label{beam_gain}
\end{aligned}
\end{equation}
To construct the polar-domain codebook $\mathbf{A}_{\mathrm{T}}(r,\theta)\in \mathbb{C}^{N \times G_{\mathrm{T}}}$, where $G_{\mathrm{T}} = \kappa_\theta \times \kappa_r$ denotes the number of codebook vectors, the angular domain is first discretized into $\kappa_\theta$ bins while keeping the distance fixed. Subsequently, the distance domain is sampled into $\kappa_r$ bins for a fixed angle. This process is described next.

\begin{figure*}
\centering
\subfloat[]{\includegraphics[scale=0.55]{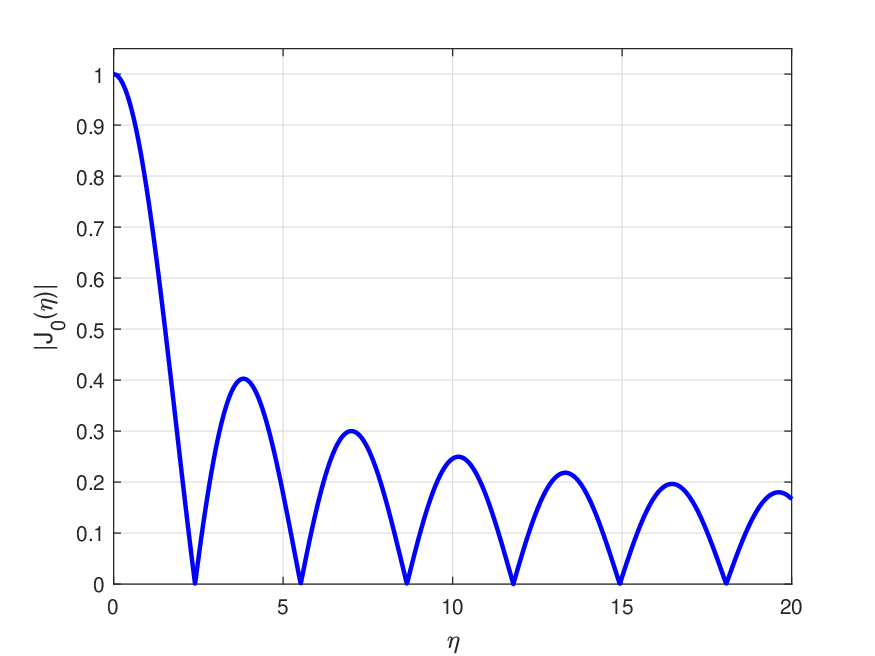}}
 \hfil
\hspace{-10pt} \subfloat[]{\includegraphics[scale=0.35]{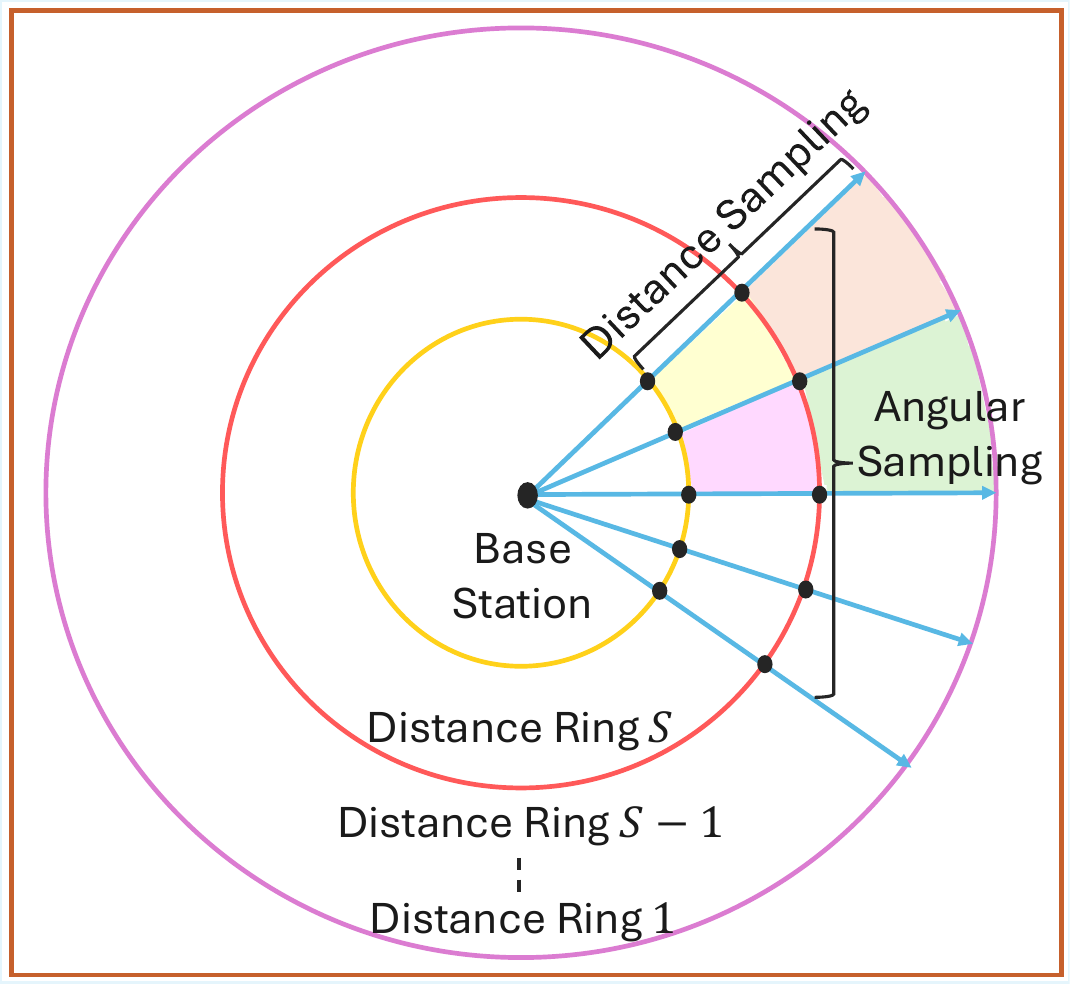}}
\caption{$ \left(a\right) $ Illustration of the magnitude of the zeroth-order Bessel function $ \left(b\right) $ Concentric ring-based polar-domain codebook illustration for UCA}
\label{besl}
\end{figure*}
\subsubsection{Angle-Domain Sampling}
Consider the case of $r_1 = r_2 = r$ in Eq~\eqref{beam_gain}. The beamfocusing gain 
$g(r,\theta_1,\theta_2)$ for this case can be approximated as the usual far-field beamforming gain (see Appendix-\ref{ang_sampling}):
\begin{align}\label{angular}
\!\!\!\!\!g(r,\theta_1,\theta_2)&=|\tilde{\mathbf{a}}^{H}(r,\theta_{1})\tilde{\mathbf{a}}(r,\theta_{2})| \\ \nonumber
    &\approx \frac{1}{N} \Big|\sum_{n=1}^{N} e^{j\eta \sin \big(\psi_n-\frac{\theta_1+\theta_2}{2}\big)}\Big|\approx |J_0({\eta})|,
\end{align}
where $\eta = \frac{4\pi R}{\lambda} \sin(\frac{\theta_{2} - \theta_{1}}{2})$ and $J_{0}(\cdot)$ denotes the zeroth-order Bessel function of the first-kind. It can be readily observed that the beamfocusing gain becomes independent of the distance $r$. Furthermore, from Fig~\ref{besl}(a), we note that the beamfocusing gain $|J_{0}(\eta)|$ attains its maximum value at $\eta = 0$. As 
$\eta$ increases, the gain exhibits a decaying behavior accompanied by oscillations. Notably, within the interval $\eta \in [0, 2.4]$, the function decreases monotonically, which is advantageous for angular-domain sampling. Based on this property, the angular separation between two neighboring codebook vectors can be selected such that $\eta \approx 2.4$ corresponding to the first zero-crossing of $|J_{0}(\eta)|$. This choice minimizes the correlation between the adjacent codewords. Following the above statement and Eq.~\eqref{angular}, we solve for $g(r, \theta_1, \theta_2) \approx |J_0(\eta)| \leq \Delta$, where $\Delta$ is defined as the threshold for maximum allowable correlation. After simplification, we obtain
\begin{align}
\eta &= \frac{4\pi R}{\lambda} \sin^{-1}\!\Big( \frac{\theta_{\Delta}}{2} \Big) 
\geq J_0^{-1}(\Delta), \nonumber \\
\Rightarrow \theta_{\Delta} &\geq 2 \sin\!\Big( \frac{\lambda}{4\pi R} J_0^{-1}(\Delta) \Big).
\end{align}
Thus, the angle-domain samples are defined as
$\Big\{ \theta_s = s \theta_{\Delta} \Big\}_{s=0}^{\kappa_\theta - 1}$, where $\kappa_\theta = \Big\lfloor \frac{2\pi}{\theta_{\Delta}}\Big\rfloor$.
\begin{algorithm}[t]\label{Al_1}
\caption{Design of a polar-domain concentric ring  codebook for near-field sub-THz MIMO systems}
\KwIn{Fresnel distance $(r_{\min})$, threshold $(\Delta)$, radius of UCA $(R)$, wavelength  $(\lambda)$} 
\KwOut{Polar codebook $\mathbf{A}_\mathrm{T}(r,\theta) \in \mathbb{C}^{N \times G_\mathrm{T}}$, where $G_\mathrm{T} = \kappa_\theta \kappa_r$
}

{
$\theta_{\Delta} = 2 \sin^{-1}\left(\dfrac{\lambda {J}_{0}^{-1}(\Delta)}{4 \pi R}\right)$, $r_{\Delta} = \dfrac{\pi R^{2}}{2 \lambda {J}_{0}^{-1}(\Delta)}$


$\kappa_\theta = \left\lfloor \frac{2\pi}{\theta_{\Delta}}\right\rfloor, \kappa_r = \left\lfloor\frac{2 \pi R^{2}}{\lambda J_{0}^{-1}(\Delta)r_{\min}} \right\rfloor$

Construct $\Big\{ r_{s_1} = \frac{r_{\Delta}}{s_1} \Big\}_{s_1=0}^{\kappa_r-1}$

Construct $\Big\{ \theta_{s_2} = s_2 \theta_{\Delta} \Big\}_{s_2=0}^{\kappa_\theta - 1}$
   
Construct $\mathbf{A}_\mathrm{T}(r,\theta)=\Big[ \mathbf{a}\Big(r_{s_{1}},  \theta_{s_2}\Big) \Big]_{\forall s_1, s_2}$ }
\end{algorithm}
\subsubsection{Distance-Domain Sampling}
To study the variation of beamfocusing gain with respect to distance, we set $\theta_{1}=\theta_{2}=\theta$ in Eq~\eqref{beam_gain}. For this case, the beamfocusing gain $g(r_{1}, \theta, r_{2})$ is approximated as (see Appendix-\ref{dist_domain_sampling}):
\begin{align}\label{distance}
g(r_{1}, \theta, r_{2}) &=|\tilde{\mathbf{a}}^{H}(r_{1},\theta)\tilde{\mathbf{a}}(r_{2},\theta)| \notag \\ 
&\approx\frac{1}{N} \left\vert \sum_{n=1}^{N} e^{j \frac{2\pi}{\lambda} \left( \Xi_{r_1,\theta}^{(n)} - \Xi_{r_2,\theta}^{(n)} \right)} \right\vert \\ 
&\approx |J_{0}(\varrho)|,\label{dis}
\end{align}
where $\varrho = \frac{2 \pi R^{2}}{\lambda} 
\big( \frac{1}{4r_{1}} - \frac{1}{4r_{2}} \big)$, and  $\Xi_{r_1,\theta}^{(n)}$, $\Xi_{r_2,\theta}^{(n)}$ are defined in Eq.~\eqref{radius}. We once again observe that the beamfocusing gain $g(r_{1}, \theta, r_{2}) \approx |J_{0}(\varrho)|$ does not depend on the angle $\theta$. Hence, in order to obtain the distance samples, we follow $|J_0(\varrho)| \leq \Delta$. Solving this criteria further gives
\begin{equation}
\begin{aligned}
   \varrho &=  \frac{2\pi R^2}{\lambda} 
\Big| 
\frac{1}{4 r_a} - \frac{1}{4 r_b} 
\Big| \geq J_0^{-1}(\Delta) \\ \nonumber
&\Rightarrow \Big| 
\frac{1}{r_a} - \frac{1}{r_b} 
\Big| 
\geq \frac{2\lambda}{\pi R^2} J_0^{-1}(\Delta). 
\end{aligned}  
\end{equation}
Therefore, the distance samples, as depicted in Fig. \ref{besl}(b), are obtained as
\begin{equation}\label{distance_sample}
r_{s} = \frac{r_{\Delta}}{s}, \quad s = 0, 1, \ldots, \kappa_r-1,
\end{equation}
where $r_{\Delta} = \dfrac{\pi R^{2}}{2 \lambda {J}_{0}^{-1}(\Delta)}$, $\kappa_r = \left\lfloor\frac{2 \pi R^{2}}{\lambda_{k} J_{0}^{-1}(\Delta)r_{\min}} \right\rfloor$. Here, $r_\mathrm{min}$ denotes the Fresnel distance and it is given as $r_\mathrm{min} = \frac{D^{4/3}}{2 \lambda^{1/3}}$ with $D=2R$ representing the UCA aperture. A pseudo code for the codebook design based on the above polar-domain sampling is presented in Algorithm-\ref{Al_1}. 

\subsection{Hybrid Precoder Design for Near-Field sub-THz MIMO Systems}
A hybrid precoder design problem for the near-field sub-THz MU-MIMO downlink system, aimed at achieving the closest possible approximation to the corresponding FD beamformer $\mathbf{W}$, can be formulated as
\begin{align}
\mathcal{P}_{3}: \min_{\{\mathbf{F}_{\mathrm{RF}}, \mathbf{F}_{\mathrm{BB}}\}} \quad & \|\mathbf{W}-\mathbf{F}_{\mathrm{RF}}\mathbf{F}_{\mathrm{BB}}\|_{F}  \label{object}
\\ \text{s.t.} &\parallel \mathbf{F}_{\mathrm{RF}} \mathbf{F}_{\mathrm{BB}} \parallel = \left \Vert \mathbf{W} \right \Vert, \label{constraint_power}\\
\quad & |\mathbf{F}_{\mathrm{RF}}(i,j)| = \frac{1}{\sqrt{N}},\label{Modul1}
\end{align}
where the constraint \eqref{constraint_power} ensures that the transmit power with hybrid precoder is same as that with the FD precoder, and \eqref{Modul1} is the usual constant modulus constraint on the elements of the RF precoder $\mathbf{F}_{\mathrm{RF}}$.
A careful examination of \eqref{stage_1} reveals that for sufficiently small values of the penalty parameter $\zeta$ 
$$\zeta \left[ 2\mathbf{I} + \zeta \sum_{j=1}^{K} \mathbf{h}_j \mathbf{h}_j^{H} \right]^{-1} \longrightarrow \frac{\zeta}{2}\mathbf{I}.$$
Consequently, the resultant digital beamformer $\mathbf{w}_{k}$ significantly depends on the quantity $\left( \sum_{j=1}^{K}\mathbf{h}_{j}c_{j,k} \right)$, which implies that the FD beamformer $\mathbf{w}_{k}$ lies in the column-space $\mathcal{C}(\mathbf{H})$ of the channel matrix $\mathbf{H}$. 

Revisiting Eq.~\eqref{channel_model}, one can rewrite the near-field channel $\mathbf{h}_k = \sqrt\frac{N}{L} \bar{\mathbf{A}}_{\mathrm{T} }\bar{\boldsymbol{\alpha}}_k$, where $\bar{\mathbf{A}}_{\mathrm{T}}\in \mathbb{C}^{N \times L}$ comprises the array steering vectors $\tilde{\mathbf{a}}(r_l,\theta_l), 1\leq l \leq L$, for the $L$ dominant multipath components, and the vector $\bar{\boldsymbol{\alpha}}_k \in \mathbb{C}^{L \times 1}$ contains their complex path gains. Thus,  we can conclude that
\begin{align}
\mathbf{w}_{k} \subseteq \mathcal{C}(\mathbf{H}) \subseteq \mathcal{C}(\bar{\mathbf{A}}_{\mathrm{T}}).
\end{align}
The above analysis, considering the objective function \eqref{object} of $\mathcal{P}_{3}$, recommends that  $\bar{\mathbf{A}}_{\mathrm{T}}$ is an appropriate choice for designing the RF precoder $\mathbf{F}_{\mathrm{RF}}$, since $\mathcal{C}(\mathbf{W}) \subseteq \mathcal{C}(\bar{\mathbf{A}}_{\mathrm{T}})$. However, since $\bar{\mathbf{A}}_{\mathrm{T}}$ is typically unknown, we replace it with the corresponding dictionary $\mathbf{A}_\mathrm{T}(r,\theta)$ designed in {Algorithm-\ref{Al_1}}. Hence, the modified hybrid precoder design problem is formulated as follows.
\begin{align}
\mathcal{P}_{4}: \min_{ \left\{ \tilde{\mathbf{F}}_{\mathrm{BB}}\right\} } \quad &  \|\mathbf{W} - \mathbf{A}_\mathrm{T}(r,\theta) \tilde{\mathbf{F}}_{\mathrm{BB}} \|_{F}  \label{sparse_MMV_prob_obj}
\\ \text{s.t.} \quad & \left\Vert \mathrm{diag}\left( \tilde{\mathbf{F}}_{\mathrm{BB}} \tilde{\mathbf{F}}_{\mathrm{BB}}^H \right) \right\Vert_0 = N_{\mathrm{RF}}, \label{sparse_MMV_prob_const} \\
&\parallel \mathbf{A}_\mathrm{T}(r,\theta) \tilde{\mathbf{F}}_{\mathrm{BB}} \parallel = \left \Vert \mathbf{W} \right \Vert, \label{tx_power_constr}
\end{align}
where $\tilde{\mathbf{F}}_{\mathrm{BB}} \in \mathbb{C}^{G_{\mathrm{T}} \times K}$, and the constraint \eqref{sparse_MMV_prob_const} is the sparsity constraint which represents that $\tilde{\mathbf{F}}_{\mathrm{BB}}$ can have only $N_{\mathrm{RF}}$ non-zero rows. This additional constraint is introduced since the final RF precoder $\mathbf{F}_{\mathrm{RF}}$ can have only $N_{\mathrm{RF}}$ columns, which implies that we can select only $N_{\mathrm{RF}}$ active columns of $\mathbf{A}_\mathrm{T}(r,\theta)$ to form the RF precoder. Accordingly, the matrix $\tilde{\mathbf{F}}_{\mathrm{BB}}$ becomes simultaneous sparse in nature with only $N_{\mathrm{RF}}$ non-zero rows, and the optimization problem $\mathcal{P}_{4}$ is an MMV-based sparse signal recovery problem, which can be efficiently solved by SOMP algorithm as outlined in Algorithm~(\ref{Al_2}). The step-wise description of this algorithm is briefly presented next.

The residue $\mathbf{F}_{\mathrm{res}}$ and RF precoder $\mathbf{F}_{\mathrm{RF}}$ are initialized with the FD precoder $\mathbf{W}$ and a null matrix, respectively. In step-3 of each iteration, we compute  projection of the residue $\mathbf{F}_{\mathrm{res}}$ onto the dictionary $\mathbf{A}_{\mathrm{T}}(r,\theta)$, which signifies the correlation between each dictionary atom and the current reside. Step-4 involves locating the column index $t$ that yields the highest correlation value, which in turn determines the beamfocusing vector that is most aligned with the current residue, and thereby it participates in constructing $\mathbf{F}_{\mathrm{RF}}$ in step-5. In step-6, the baseband precoder $\mathbf{F}_{\mathrm{BB}}$ is obtained via a least-squares solution, after which the residue is subsequently updated in step-7. Hence, after running $N_{\mathrm{RF}}$ iterations, we yield the RF precoder $\mathbf{F}_{\mathrm{RF}}$ of dimension $N \times N_{\mathrm{RF}}$ and the baseband precoder $\mathbf{F}_{\mathrm{BB}}$ of dimension $N_{\mathrm{RF}} \times K$. In the last, step-8 normalizes  $\mathbf{F}_{\mathrm{BB}}$ to satisfy the transmit power constraint of \eqref{constraint_power}.
\begin{algorithm}[t]\label{Al_2}
\caption{Near-field hybrid precoder design via SOMP}
\KwIn{Equivalent digital beamformer $(\mathbf{W})$, near-field polar-domain concentric-ring codebook ${\mathbf{A}}_{\mathrm{T}} = {\mathbf{A}}_{\mathrm{T}}(r,\theta)$, number of RF chains $N_{\mathrm{RF}}$}
\KwOut{RF precoder $\mathbf{F}_\mathrm{RF}$, baseband precoder $\mathbf{F}_\mathrm{BB}$
}

\textbf{Initialize:} $\mathbf{F}_{\mathrm{RF}}=[\emptyset]$, $\mathbf{F}_{\mathrm{res}}=\mathbf{W}$  \\
\For{$j=1:N_{\mathrm{RF}}$}{
$\mathbf{\Psi} = {\mathbf{A}}^H_{\mathrm{T}} \mathbf{F}_{\mathrm{res}}$

$t=\mathrm{arg max} \ \  \mathrm{diag}(\mathbf{\Psi}\mathbf{\Psi}^{H})$

Construct $\mathbf{F}_{\mathrm{RF}}=[ \mathbf{F}_{\mathrm{RF}} \ | \ {\mathbf{A}}_{\mathrm{T}}(:,t) ]$

Construct $\mathbf{F}_{\mathrm{BB}}=\left( \mathbf{F}^{H}_{\mathrm{RF}} \mathbf{F}_{\mathrm{RF}}\right)^{-1}\mathbf{F}^{H}_{\mathrm{RF}}\mathbf{W}$

Construct $\mathbf{F}_{\mathrm{res}}=\frac{\mathbf{W}-\mathbf{F}_{\mathbf{RF}}\mathbf{F}_{\mathrm{BB}}}{\|\mathbf{W}-\mathbf{F}_{\mathrm{RF}}\mathbf{F}_{\mathrm{BB}}\|_{F}}$ }
{$\mathbf{F}_{\mathrm{BB}}=\frac{\mathbf{F}_{\mathrm{BB}}\left \Vert \mathbf{W} \right \Vert_{F}}{\| \mathbf{F}_{\mathrm{RF}} \mathbf{F}_{\mathrm{BB}} \|_{F}}$}
\end{algorithm}

\section{Simulation Results}\label{simulation}
The performance of the proposed PDJO framework for near-field sub-THz MU beamforming is presented next in terms of transmit power utilized, sum SE and system EE w.r.t. the target SINR $\gamma_k = \gamma, \forall k$. To quantify performance, Table~\ref{system_para} outlines our simulation parameters. The AoDs $\theta_{i}$ are chosen uniformly within the interval~$[-\pi,\pi)$. The auxiliary variables $\{c_{k,j}  \}$ and penalty factor $\zeta$ are initialized with $c_{k,j} \sim \mathcal{CN}(0,1)$ and $\zeta=10^{-3}$, whereas the constant $p$ is set to $0.9$ for updating the penalty factor $\zeta$. The correlation threshold $\Delta$ for polar-domain codebook design of {Algorithm-\ref{Al_1}} is set to $0.4$ and the convergence parameter $\epsilon$ of {Algorithm-\ref{penalty_algorithm}} is set to $10^{-7}$. 

\begin{table}
\centering
\caption{Simulation parameters} 
\label{system_para}
\resizebox{0.42\textwidth}{!}{%
\begin{tabular}{|l|c|c|}\hline
\textbf{Parameter} & \textbf{Symbol} & \textbf{Value}  \\\hline
$\#$ of BS antennas & $N$ &  $128$  \\\hline
BS UCA radius & $R$ & $0.13$m  \\\hline
$\#$ of BS RF chains & $N_{\mathrm{RF}}$ &  $8$  \\\hline
$\#$ of users & $K$ &  $4$  \\\hline
$\#$ of scatterers & $L$ &  $5$ \\\hline
BS grid size  & $G_{\mathrm{R}} =(\kappa_\theta \times \kappa_r )$  &  $(1084 \times 9)$  \\\hline
Operating frequency & $f_c$ &  $100$\:GHz  \\\hline
Noise power at users & $\sigma^2_{k}$ &  $1$  \\\hline
Power amplifier efficiency & $\mu$ &  $0.3$\cite{zou2022optimal}  \\\hline
Circuit power & $P_{c}$ &  $30$ dBm\cite{zou2022optimal}  \\\hline
\end{tabular}%
}
\end{table}

\begin{figure*} 
	\centering
	\subfloat[]{\includegraphics[scale=0.42]{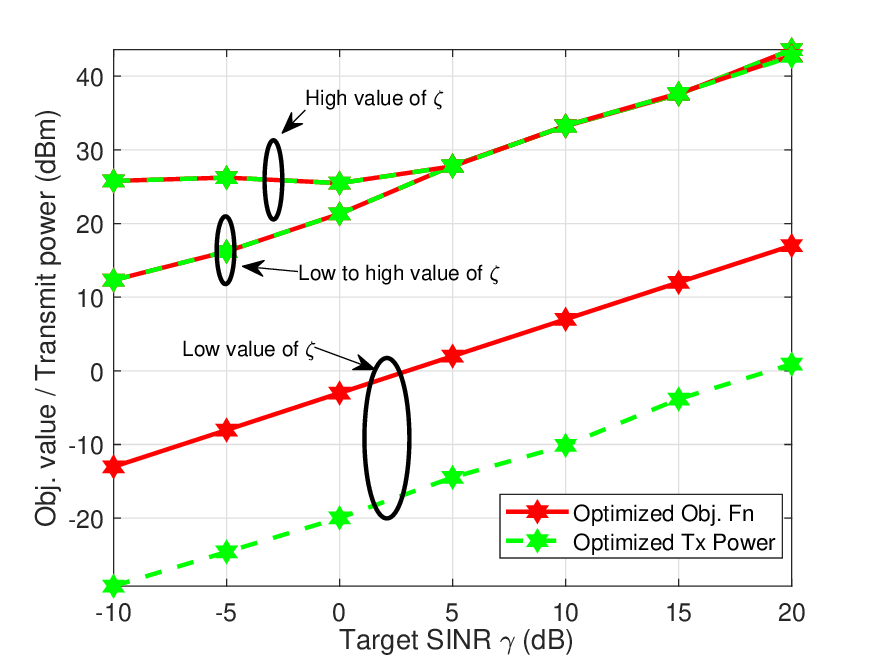}}
	\hfil
	\hspace{-12pt}\subfloat[]{\includegraphics[scale=0.42]{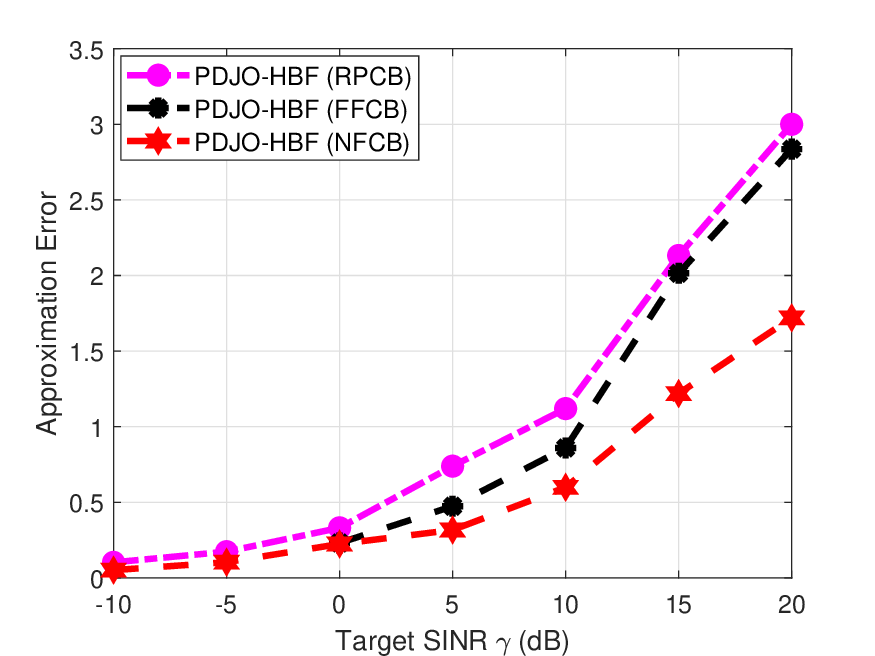}}
 	\hfil
	\hspace{-12pt} \subfloat[]{\includegraphics[scale=0.42]{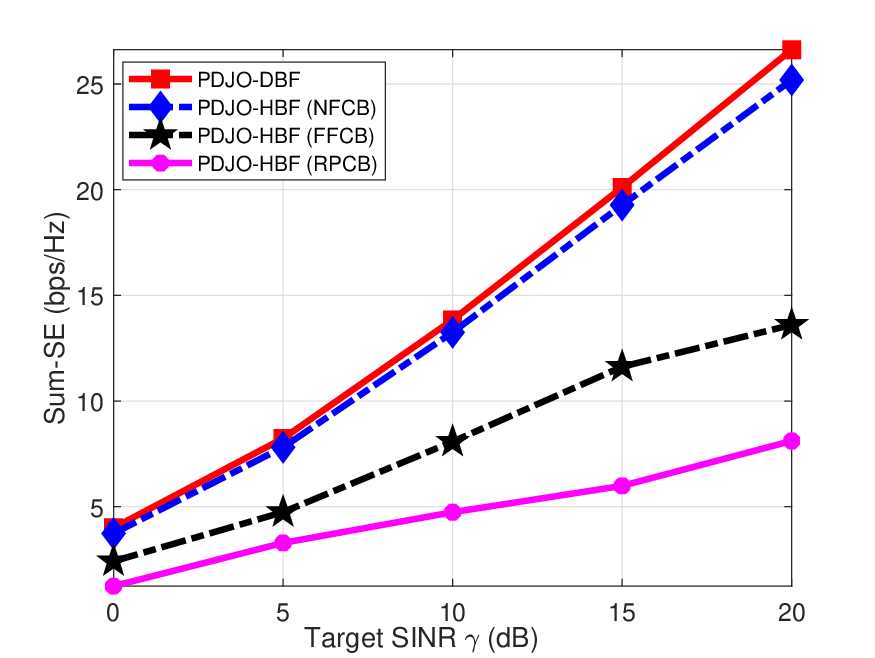}}
    \caption{$ \left(a\right) $ Transmit power versus target SINR at different value of penalty parameter $\zeta$; $ \left(b\right) $ Approximation error versus target SINR considering near-field codebook (NFCB), far-field codebook (FFCB) and random-precoding codebook (RPCB); $ \left(c\right) $ Sum-SE in (bits/sec/Hz) versus target SINR for the proposed digital beamfocusing (DBF) and hybrid beamfocusing (HBF) architectures.}
	\label{results} 
\end{figure*}

Fig. \ref{results}(a) illustrates transmit power utilized in dBm by the hybrid precoder designed w.r.t. target SINR~$\gamma$ in dB. To achieve these results, first {Algorithm-\ref{penalty_algorithm}} is run for designing a FD MU beamformer $\mathbf{W}$, followed by decomposing it into the corresponding hybrid precoder (RF precoder $\mathbf{F}_\mathrm{RF}$, baseband precoder $\mathbf{F}_\mathrm{BB}$) using {Algorithm-\ref{Al_2}}, which in turn utilizes the polar-domain codebook $\mathbf{A}_\mathrm{T}(r,\theta)$ designed in {Algorithm-\ref{Al_1}}. Finally, the transmit power $\parallel \mathbf{F}_{\mathrm{RF}}\mathbf{F}_{\mathrm{BB}} \parallel^2$ has been plotted w.r.t. the target SINR $\gamma$. We observe that the transmit power increases upon increasing the target SINR threshold to satisfy users' QoS requirements. Furthermore, for a low value of the penalty parameter $\zeta=0.001$, the objective function defined in~(\ref{problem_1}) and the transmit power utilized have significant gap at all values for target SINR~$\gamma$. This is because the first term of the objective function~(\ref{problem_1}) dominates for low value of $\zeta$ and the constraints $\mathbf{h}^H_{k} \mathbf{w}_{j}=c_{k,j}$ may not be strictly satisfied. Therefore, the optimization primarily focuses on minimizing the transmit power in unconstrained manner. On the other hand, for a large value of the penalty parameter {$\zeta=100$}, the objective function value and the transmit power strictly coincide owing to strongly enforcing the constraints $\mathbf{h}^H_{k}\mathbf{w}_{j} \rightarrow   c_{k,j}$. However, for this case, the transmit power utilized is on a higher side as compared to the scenario when we set $\zeta=0.001$. Therefore, we begin with a small value of $\zeta$ to prioritize transmit power minimization and to obtain a suitable point for initialization, and then progressively increase $\zeta$, as specified in step-7 of {Algorithm-\ref{penalty_algorithm}}, so that the constraint $\mathbf{h}^H_{k}\mathbf{w}_{j} \rightarrow  c_{k,j}$ is enforced while maintaining a balance between objective function and transmit power. The related results encircled as \textit{low to high value of $\zeta$} demonstrate that the constraints are satisfied and the transmit power is also lower as compared to \textit{high value of $\zeta$} case, especially at lower target SINR values. This justifies our insights developed regarding the proposed PDJO framework toward the end of  Section-\ref{PD}. 

\begin{figure*} 
	\centering
	\subfloat[]{\includegraphics[scale=0.42]{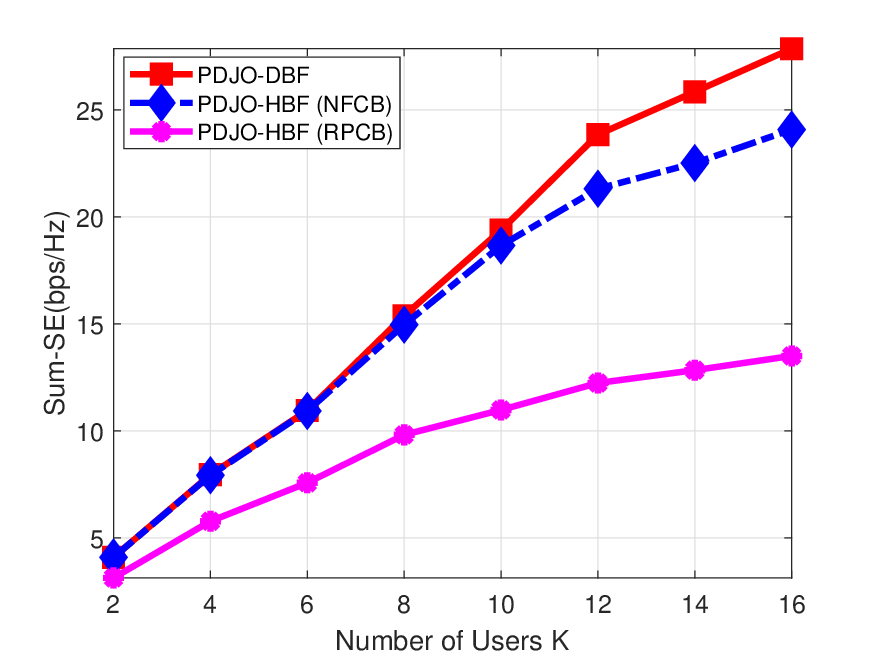}}
	\hfil
	\hspace{-12pt}\subfloat[]{\includegraphics[scale=0.42]{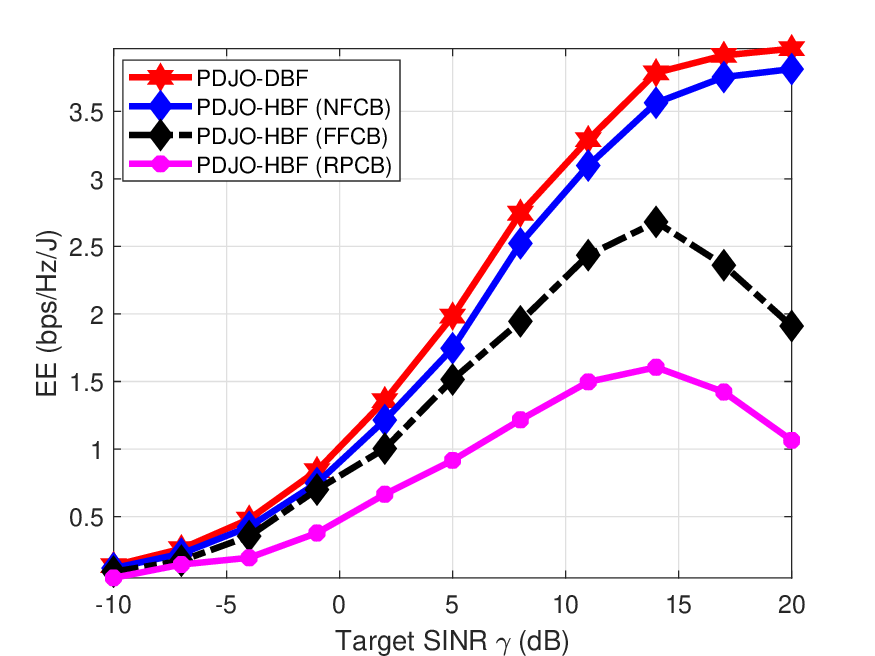}}
 	\hfil
	\hspace{-12pt} \subfloat[]{\includegraphics[scale=0.42]{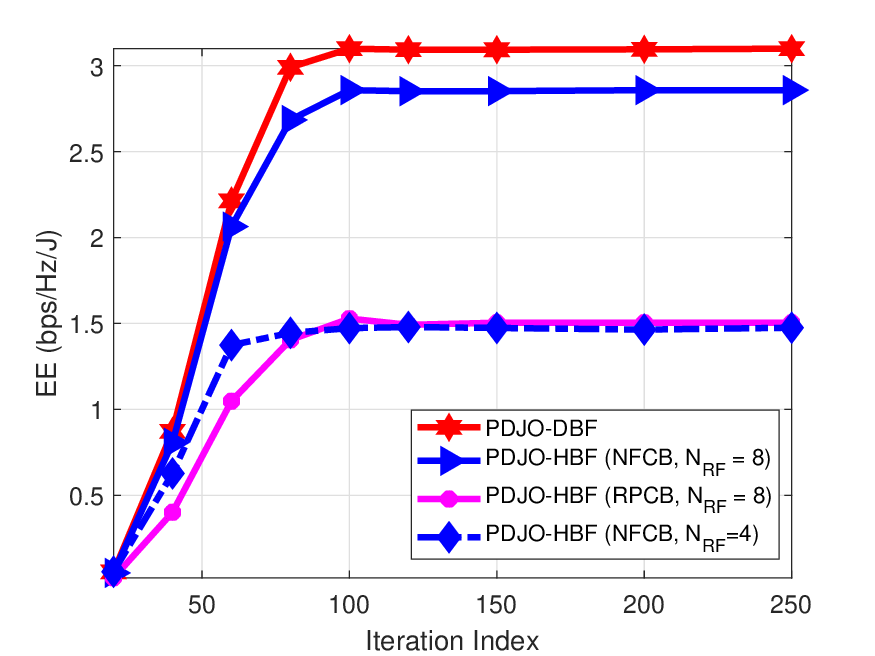}}
    \caption{$ \left(a\right) $ Sum-SE vs number of Users (K) in near-field; $ \left(b\right) $ EE vs target SINR ($\gamma$); $ \left(c\right) $ EE vs Iteration index at $\gamma=10$ dB considering near-field and randomized codebook.}
	\label{results2} 
\end{figure*}

Fig.~\ref{results}(b) illustrates the variation of the approximation error with respect to the target SINR $\gamma$. The approximation error, defined as
$\|\mathbf{W}-\mathbf{F}_{\mathrm{RF}} \mathbf{F}_{\mathrm{BB}}\|$, quantifies the error in decomposing the FD beamformer $\mathbf{W}$ into the corresponding hybrid beamformer $\mathbf{F}_{\mathrm{RF}}$ and $\mathbf{F}_{\mathrm{BB}}$. From the plot, it can be observed that the approximation error increases as the target SINR $\gamma$ grows. We notice this trend because achieving higher SINR levels requires more stringent beamfocusing accuracy and therefore increased transmit power. Furthermore, we note that modeling a near-field channel using a far-field codebook (FFCB) leads to a large mismatch in approximating the FD precoder, resulting in increasing approximation error w.r.t. that of the near-field codebook (NFCB). This result also considers a randomized RF precoding using RPCB, where the RF precoder is configured using random phase shifts and its elements are given as  $\mathbf{F}_{\mathrm{RF}}(\imath,\jmath)
= \frac{1}{\sqrt{N}} \exp\left(j\omega_{\imath,\jmath}\right)$. Here the phase values $\omega_{\imath,\jmath}$ are chosen uniformly randomly from the discrete set $
{\Upsilon} = \left\{ 0, \frac{2\pi}{2^{N_Q}}, \ldots, \frac{(2^{N_Q}-1)2\pi}{2^{N_Q}} \right\}$ with $N_{Q}=6$-bit phase shifters. 


Fig.~\ref{results}(c) shows the sum-SE, i.e., $\sum_{k=1}^{K}\log_{2}(1+\text{SINR}_{k}),$ w.r.t. target SINR for different beamfocusing schemes. For this, the FD  beamfocusing (DBF) obtained via the proposed PDJO framework is considered as a benchmark. This is due to the fact that DBF employs a dedicated RF chain for each antenna element, i.e., $N_{\mathrm{RF}} = N$, which enhances beamfocusing flexibility at the expense of increased hardware cost and system complexity. By contrast, hybrid beamfocusing (HBF) architecture utilizes a significantly reduced number of RF chains at the cost of a slight performance degradation. The sum-SE is evaluated considering the HBF using both the proposed NFCB of {Algorithm-\ref{Al_1}} and the usual FFCB \cite{heath2016overview}. It can be readily observed that the sum-SE obtained from the proposed method closely approaches that of the DBF architecture. On the other hand, the same degrades significantly when a conventional FFCB is used for HBF. This is because such a FFCB fails to accurately capture the near-field propagation characteristics. Also, the RPCB is seen to perform worst due to a large approximation error.

The sum-SE as a function of number of users $K$ at a fixed target SINR $\gamma_{k} = 5$ dB is illustrated in Fig.~\ref{results2}(a). We observe that the sum-SE initially increases with number of users due to enhanced spatial multiplexing gains. However, after $K=12$, the sum-SE growth rate diminishes, primarily due to the increased impact of MU interference. Furthermore, for the HBF architecture, proposed beamfocusing scheme achieves performance comparable to that of the DBF system for a moderate number of users. Specifically, with $N_{\mathrm{RF}}=8$ RF chains, the sum-SE of the HBF architecture closely matches to that of the DBF architecture up to $8$ users, and then starts deviating. As the number of users increases further, the limited number of RF chains restricts the ability of the HBF system to approximate the DBF system, and in turn to efficiently serve all users. As a result, a significant performance gap emerges between  DBF and HBF at higher user densities. However, this gap can be filled by further increasing the number of RF chains $N_{\mathrm{RF}}$.

Fig.~\ref{results2}(b) illustrates the variation of EE w.r.t. target SINR which is computed as  
\begin{align}
    \text{EE} =\frac{\sum_{k=1}^K r_{k}}{\mu \sum_{k=1}^{K} \left\Vert \mathbf{w}_{k} \right\Vert^2+N_{\mathrm{RF}}P_{c}},
\end{align}
where $r_{k}=\log_{2}(1+\text{SINR}_{k})$ denotes the $k$-th user's achievable rate, $\mu$ represents the power amplifier efficiency and $P_{c}$ is the circuit power. EE refers to how efficiently a system utilizes energy to transmit the overall data rate. From the plot, it is observed that for both PDJO-DBF and PDJO-HBF utilizing NFCB, the EE initially increases as target SINR increases, attains a maximum value, and then it gets saturated with further increases in SINR. This behavior is seen due to  increase in transmit power and achievable sum-rate upon increasing the target SINR, thereby improving the EE. However, for PDJO-HBF with FFCB, increase in transmit power outweighs the rate improvement, leading to a decline in EE after SINR of $13$ dB. Furthermore, we crucially observe that the EE of the proposed PDJO-HBF with NFCB closely follows that of the PDJO-DBF.

Finally, Fig~\ref{results2}(c) illustrates the convergence behavior of the proposed PDJO-based methods by plotting EE versus number of iterations for different values of RF chains $N_{\mathrm{RF}}$. We observe that the EE attains its peak value and converges within approximately 100 iterations for both PDJO-DBF and PDJO-HBF with $N_{\mathrm{RF}} = 8$. A similar convergence behavior is observed for $N_{\mathrm{RF}} = 4$; however EE is seen to reduce significantly for this setting. 

\section{Conclusion}\label{conclusion} This paper investigated the design of energy-efficient beamfocusing strategies for near-field sub-THz MU-MIMO communications with particular emphasis on hybrid precoder design employing a large UCA at the BS. To this end, a QoS-aware optimization framework was formulated to minimize the BS transmit power while satisfying the individual SINR requirements of all users. The resulting non-convex optimization problem was efficiently addressed through the proposed two-stage penalty-driven joint optimization (PDJO) framework. Furthermore, a near-field polar-domain codebook was developed to accurately characterize spherical wavefront propagation, enabling the decomposition of the FD precoder into hybrid RF and baseband precoders using the proposed SOMP-based sparse approximation framework.

Simulation results demonstrated that the proposed framework effectively balances transmit power minimization and QoS satisfaction through appropriate selection of the penalty parameter. In addition, the proposed near-field polar-domain codebook consistently outperformed conventional far-field codebooks, highlighting the importance of accurately accounting for spherical wavefront propagation in near-field beamfocusing. Finally, the proposed hybrid beamfocusing framework achieved a favorable trade-off between hardware complexity and system performance, making it a promising solution for energy-efficient sub-THz near-field MU-MIMO communications.

\appendices

\section{mathematical derivation of stage-1 beamfocuser design} \label{PDJO_derivation}
Taking the gradient of \eqref{opt} w.r.t. $\mathbf{w}_{k}$  and equating it to zero, we obtain
\begin{align}
2\mathbf{w}_k
+\zeta \sum_{j=1}^{K} \mathbf{h}_j \mathbf{h}_j^H \mathbf{w}_k
-\zeta \sum_{j=1}^{K} \mathbf{h}_j c_{j,k}
= 0.
\end{align}
Upon rearranging the terms, the resultant expression can be written as follows:
\begin{align}
\left(
2\mathbf{I}+\zeta \sum_{j=1}^{K} \mathbf{h}_j \mathbf{h}_j^H
\right) \mathbf{w}_k
&=
\zeta \sum_{j=1}^{K} \mathbf{h}_j c_{j,k}.
\end{align}
Further solving it for $\mathbf{w}_{k}$ yields
\begin{align}
 \mathbf{w}_k
&=
\zeta \left(
2\mathbf{I}+\zeta \sum_{j=1}^{K} \mathbf{h}_j \mathbf{h}_j^H
\right)^{-1}\sum_{j=1}^{K} \mathbf{h}_j c_{j,k},
\end{align}
which completes the proof of ideal FD beamfocusing design.

\section{mathematical approximations of angular-domain sampling} \label{ang_sampling}
Assuming $r_{1}=r_{2}=r$, and substituting \eqref{radius} into the beamfocusing expression $g(r_{1},\theta_{1},r_2,\theta_{2})=\big|\mathbf{a}^H(r_1,\theta_1)\mathbf{a}(r_2,\theta_2)\big|$, we obtain
\begin{align*} 
g(r, \theta _{1}, r, \theta _{2}) 
& = \frac {1}{N} \left\vert { \sum _{n=1}^{N} e^{j\frac {2\pi }{\lambda }\sqrt {r^{2}+R^{2}-2rR\cos (\theta _{1} - \psi _{n})}}}\Big. \Big.{ \times e^{-j\frac {2\pi }{\lambda }\sqrt {r^{2}+R^{2}-2rR\cos (\theta _{2} - \psi _{n})}} \vphantom {\left |{ \sum _{n=1}^{N} e^{j\frac {2\pi }{\lambda }\sqrt {r^{2}+R^{2}-2rR\cos (\theta _{1} - \psi _{n})}}}\right.}}\right\vert \\ 
& \mathop {\approx }\limits ^{(a)} \frac {1}{N} \Big |{ \sum _{n=1}^{N} e^{-j\frac {2\pi }{\lambda }R \left [{\cos (\theta _{1} - \psi _{n}) - \cos (\theta _{2} - \psi _{n})}\right]} }\Big | . 
\end{align*}
The approximation in (a) is obtained by employing the Taylor series expansion $\sqrt{1 + x} = 1 + \frac{x}{2} - \frac{x^2}{8} + \mathcal{O}(x^3)$ and retaining only the first-order term. Under this linear approximation, the propagation distance $r(n)$ reduces to its far-field (angular-domain) form. By further simplifying using trigonometric identity: $\cos(C)-\cos(D) = -2 \sin(\frac{C+D}{2}) \sin(\frac{C-D}{2})$, the above expression can be reformulated as
\begin{align} \label{trignometric}
g(r, \theta _{1}, r, \theta _{2}) &\mathop {=} \frac {1}{N} \Big |{ \sum _{n=1}^{N} e^{j\eta \sin \left({\psi _{n} - \frac{\theta _{1}+\theta _{2}}{2}}\right)} }\Big |,
\end{align}
where $\eta = \frac{4\pi R}{\lambda} \sin(\frac{\theta_{2} - \theta_{1}}{2})$. By using $m$-th order Jacobi-Anger expansion of Bessel function as $e^{j\eta \cos \gamma } = \sum _{m=-\infty }^{\infty } j^{m} J_{m}(\eta) e^{jm\gamma }$, the above expression can be further solved as:
\begin{align} \label{bessel2}
&g(r, \theta _{1}, r, \theta _{2}) \mathop {\approx }\limits ^{(b)} \frac {1}{N}\Big |{\sum _{n=1}^{N} \sum _{m=-\infty }^{\infty } j^{m} J_{m}(\eta) e^{jm\left({\frac {\pi }{2}-\psi _{n}+\frac {\theta _{1}+\theta _{2}}{2}}\right)} }\Big | \nonumber \\ 
&= \frac {1}{N}\Big |{\sum _{m=-\infty }^{\infty } j^{m} J_{m}(\eta) e^{jm(\frac {\theta_{1}+\theta_{2}+\pi }{2})} \sum _{n=1}^{N} e^{-jm\psi _{n}} }\Big |. 
\end{align}
The formulation in (\ref{trignometric}) is expressed using a sine function, whereas the standard definition of the Bessel function is typically associated with a cosine argument. To reconcile this difference, approximation (b) is obtained by exploiting the inherent $\pi/2$ phase-shift relationship between sine and cosine functions. We further simplify the expression given in~\eqref{bessel2} by considering two cases as:
\subsubsection{\textbf{\textit{{Case-$1$}}}}
$m={N}$ 
\begin{align*}
&\sum_{n=1}^{N} e^{-j m \psi_n} 
= \sum_{n=1}^{N} e^{-j m \left(\frac{2\pi n}{N}\right)} =N\nonumber \\
\end{align*}

\subsubsection{\textbf{\textit{{Case-$2$}}}}

 $m\neq {N}$ 
 \begin{align}
\sum_{n=1}^{N} e^{-j m \psi_n} 
= \sum_{n=1}^{N} e^{-j \frac{2\pi m}{N}n} = \frac{1 - \Big(e^{-j \frac{2\pi m}{N}}\Big)^{N}}{1 - e^{-j \frac{2\pi m}{N}}} = 0.
 \nonumber
\end{align}
Based on the above two cases, Eq.~\eqref{bessel2} can be further simplified as:
\begin{align} \label{bessel3}
\hspace {-1.2pc}g(r, \theta _{1}, r, \theta _{2}) 
&= \Big |{\sum _{m=-\infty }^{\infty } j^{m} J_{m}(\eta) e^{jm(\frac {\theta_{1}+\theta_{2}+\pi }{2})} }\Big |. 
\end{align}
To accurately analyze the beamfocusing gain, it is essential to examine the behavior of the function $J_{m}(\cdot)$. Therefore, we proceed by introducing the asymptotic characterization of the Bessel function of the $m$-th order as:
\begin{equation} 
\left |{ J_{|m|}(\eta) }\right | \leq \left ({\frac {\eta e}{2|m|} }\right) ^{|m|}.
\end{equation}
It can be noted that the above expression demonstrates a diminishing trend; specifically, the magnitude $|J_{m}(\eta)|$ approaches zero when $m=N$ with large $N$. Consequently, the contribution of higher-order terms becomes negligible. Hence, \eqref{bessel3} can be accurately approximated by retaining only the dominant component associated with $m=0$ as
\begin{equation} \label{Jo}
g(r, \theta _{1}, r, \theta _{2}) \approx \left |{ J_{0}(\eta) }\right |.
\end{equation}
Up to this point, it has been established that the beamfocusing gain can be well approximated by the zeroth-order Bessel function. Furthermore, it is important to note that the accuracy of this approximation can be characterized by an associated approximation error, which quantifies the deviation between the exact expression and its simplified representation, i.e., $\delta _{g}= \left |{g(r, \theta _{1}, r, \theta _{2})-J_{0}(\eta)}\right |$. It can be further solved from Eq. \eqref{bessel2} as
\begin{align} \label{approx error}
g(r_1,\phi_1,r_2,\phi_2) &= \frac{1}{N} \Big| 
    \underbrace{J_0(\eta)N}_{m=0}
    + \underbrace{j^{N} {N} J_{N}(\eta)
        e^{j {N} \left( \frac{\theta_1 + \theta_2 + \pi}{2} \right)}}_{m={N}} 
    + \underbrace{j^{-{N}} {N} J_{-{N}}(\eta)
        e^{-j {N} \left( \frac{\theta_1 + \theta_2 + \pi}{2} \right)}}_{m=-{N}}
    \Big|\notag,
\end{align}
which implies
\begin{align}
        \big|{g(r, \theta _{1}, r, \theta _{2})-J_{0}(\eta)}\big |  \mathop {\approx }\limits ^{(c)}\Big |{ j^{N} J_{N}(\eta)\!\! \times \!\!2\cos \big({\frac {\theta _{1}+\theta _{2}+\pi }{2}}\big){N} }\Big | \leq 2\Big ({\frac {\eta e}{2{N}} }\Big)^{{N}}.
\end{align}
The approximation $(c)$ is derived by the fact: i) the symmetry property of the Bessel function, given as $J_{-{N}}(\eta) = J_{N}(\eta)$~\cite{wu2023enabling}; ii) Euler’s identity, i.e., $2\cos(\theta) = e^{j\theta} + e^{-j\theta}$. Furthermore, we crucially observe that the approximation error defined in Eq.~\eqref{approx error} exhibits a monotonic decay with respect to $N$. Consequently, as $N$ increases, the approximation becomes more precise. In particular, for sufficiently large values of $N$ the expression in Eq. \eqref{Jo} provides an accurate representation. This concludes the proof.

\section{mathematical approximations of distance-domain sampling} \label{dist_domain_sampling} 
By considering $\theta_1 = \theta_2$, the beamfocusing expression in Eq. \eqref{distance} can be rewritten as 
\begin{align}\label{approx_distance}
g(r_{1}, \theta, r_{2}, \theta) &\approx \frac{1}{N} \bigg|\sum_{n=1}^{N} 
e^{j \frac{2\pi}{\lambda} 
\left( \varrho_{r_1,\theta}^{(n)} - \varrho_{r_2,\theta}^{(n)} \right)}\bigg| \notag \\
&= \frac{1}{N} \bigg|\sum_{n=1}^{N} 
e^{j \frac{2\pi}{\lambda}
\left[
\frac{R^2}{2r_1}\left(1 - \cos^2(\theta - \psi_n)\right)
- \frac{R^2}{2r_2}\left(1 - \cos^2(\theta - \psi_n)\right)
\right]}\bigg| \notag
\end{align}
\begin{align}
    &\mathop {=}\limits ^{(d)}\frac{1}{N} \Big| 
\sum_{n=1}^{N} 
e^{j \frac{2\pi}{\lambda} 
\left( \frac{R^2}{4r_1} - \frac{R^2}{4r_2} \right)
(1 - \cos(2\theta - 2\psi_n)) }
\Big|\notag \\
&\mathop {=}\limits ^{(e)}\frac{1}{N} \Big| 
\sum_{n=1}^{N} e^{j \rho} 
e^{j \bar{\rho}
\cos(2\theta - 2\psi_n)}
\Big|.
\end{align}
In the above, equality $(d)$ is obtained by substituting the \eqref{distance} and trigonometric identity $ \cos^2(t) = \frac{1 + \cos(2t)}{2}$, which makes the quadratic cosine term into a linear combination. Furthermore, it can be observed that the parameters $\varrho = \frac{2\pi R^2}{\lambda}
\big( \frac{1}{4r_1} - \frac{1}{4r_2} \big)$ and $\bar{\varrho} = \frac{2\pi R^2}{\lambda}
\big(- \frac{1}{4r_1} + \frac{1}{4r_2} \big)$ are independent of $n$. Therefore, the above result of \eqref{approx_distance} can be simplified as
\begin{align}
g(r_1, \theta, r_2, \theta)
&=  e^{j \varrho}
\frac{1}{N}\Big|\sum_{n=1}^{N} 
e^{j \bar{\varrho} \cos(2\theta - 2\psi_n)} \Big| \notag\\
&= \frac{1}{N}
\Big|
\sum_{n=1}^{N} 
e^{j \bar{\varrho} \cos(2\theta - 2\psi_n)} 
\Big|.
\end{align}
By invoking the Jacobi–Anger expansion, as detailed in Appendix-{\ref{ang_sampling}}, the above expression can be further solved as
\begin{align}
g(r_1, \theta, r_2, \theta) 
&= \frac{1}{N} \bigg|\sum_{n=1}^{N} 
    \sum_{m=-\infty}^{\infty} 
        j^m J_m(\bar{\varrho}) 
        e^{j m (2\theta - 2\psi_n)}\bigg| \notag\\[6pt]
&= \frac{1}{N} 
    \bigg|\sum_{m=-\infty}^{\infty} 
        j^m J_m(\bar{\varrho}) 
        e^{j m 2\theta}
        \sum_{n=1}^{N} 
            e^{-j m 2\psi_n}\bigg|\notag\\
            &\mathop {\approx }\limits ^{(f)} \left |{ J_{0}(\bar{\varrho}) }\right |,
\end{align}
where the approximation $(f)$ is obtained in a similar way as for Eq.~(\ref{Jo}). Note that the Bessel function $J_{0}(\cdot)$ is an even function, which implies symmetry with respect to its argument. Consequently, this property allows us to express $J_{0}(\varrho) = J_{0}(\bar{\varrho})$. This completes the proof.

\bibliographystyle{IEEEtran}
\bibliography{physical_layer_sec.bib}
\end{document}